\documentstyle[12pt]{article}
\font\twlmsb =msbm10 scaled \magstep1
\font\egtmsb =msbm8
\font\sevmsb =msbm7
\newfam\msbfam
\textfont\msbfam\twlmsb
\scriptfont\msbfam\egtmsb
\scriptscriptfont\msbfam\sevmsb
\def\Bbb{\protect\pBbb}
\def\pBbb{\relax\ifmmode\expandafter\Bb\else\typeout{You cann't use
Bbb in text mode}\fi}
\def\Bb #1{{\fam\msbfam\relax#1}}
\def\thebibliography#1{\bigskip\section*{\centering
References\\}\bigskip\list
{\arabic{enumi}.}{\settowidth\labelwidth{#1}\leftmargin\labelwidth
\advance\leftmargin\labelsep\usecounter{enumi}}
\def\newblock{\hskip .11em plus .33em minus .07em}
\sloppy\clubpenalty4000\widowpenalty4000 \sfcode`\.=1000\relax}

\def\op#1{\mathop{\fam0 #1}\limits}

\newcommand{\Id}{{\rm Id\,}}

\newcommand{\ben}{\begin{eqnarray}}
\newcommand{\een}{\end{eqnarray}}
\newcommand{\be}{\begin{eqnarray*}}
\newcommand{\ee}{\end{eqnarray*}}
\newcommand{\bea}{\begin{eqalph}}
\newcommand{\eea}{\end{eqalph}}
\newcommand{\cL}{{\cal L}}

\newcommand{\cF}{{\cal F}}
\newcommand{\cD}{{\cal D}}
\newcommand{\al}{\alpha}
\newcommand{\bt}{\beta}
\newcommand{\dl}{\delta}
\newcommand{\la}{\lambda}

\newcommand{\g}{\gamma}
\newcommand{\G}{\Gamma}

\newcommand{\e}{\epsilon}

\newcommand{\si}{\sigma}
\newcommand{\Si}{\Sigma}

\newcommand{\f}{\phi}

\newcommand{\wt}{\widetilde}
\newcommand{\wh}{\widehat}

\newcommand{\dr}{\partial}

\newcounter{eqalph}
\newcounter{equationa}

\newenvironment{eqalph}{\stepcounter{equation}
\setcounter{equationa}{\value{equation}}
\setcounter{equation}{0}

\begin{eqnarray}}{\end{eqnarray}
\setcounter{equation}{\value{equationa}}}

\begin{document}
\hbox{}

\centerline{\large\bf GRAVITY AS A HIGGS FIELD.}
\medskip

\centerline{\large\bf III. Nongravitational Deviations}
\medskip
\centerline{\large\bf of Gravitational Fields.}
\bigskip

\centerline{\bf Gennadi A Sardanashvily}
\medskip

\centerline{Department of Theoretical Physics, Physics Faculty,}

\centerline{Moscow State University, 117234 Moscow, Russia}

\centerline{E-mail: sard@grav.phys.msu.su}

\begin{abstract}

In Parts I,II of the work (gr-qc/9405013, 9407032), we have shown that
gravity is {\it sui generis} a Higgs field corresponding to spontaneous
symmetry breaking when the fermion matter admits only the Lorentz subgroup
of world symmetries of the geometric arena. From the mathematical viewpoint,
the Higgs nature of gravity issues from the fact that different
gravitational fields are responsible for nonequivalent representations of
cotangent vectors to a world manifold by $\gamma$-matrices on spinor
bundles. It follows that gravitational fields fail to form an affine
space modelled on a linear space of deviations of some background field.
In other words, even weak gravitational fields do not
satisfy the superposition principle and, in particular, can not be quantized
by usual methods. At the same time, one can examine superposable deviations
$\sigma$ of a gravitational field $h$ so that $h+\sigma$ fail to be
a gravitational field. These deviations get the adequate mathematical
description in the framework of the affine group gauge theory in
dislocated manifolds, and their Lagrangian densities differ from the
familiar gravitational Lagrangian densities. They make contribution to the
standard gravitational effects, e.g., modify Newton's gravitational potential.
\end{abstract}

\section{Introduction}

In the naive manner, one usually describes deviations $\e$ of a
gravitational field $g$ as small deviations of ordinary physical field:
\ben
&& g'^{\mu\nu}= g^{\mu\nu} +\e^{\mu\nu},\nonumber\\
&& g'_{\mu\nu}=g_{\mu\nu}-\e_{\mu\nu}\simeq g_{\mu\al}g_{\nu\bt}\e^{\al\bt}.
\label{2.6}
\een
These deviations however fail to be superposable even in the first
order of $\e$ if one does not ignore the geometric nature of gravity
and its physical pecularity as a Higgs field.

Gravitation theory is theory with
spontaneous symmetry breaking since the fermion matter admits only the
Lorentz subgroup of world symmetries of the geometric arena.
In other words, this spontaneous symmetry
breaking appears when one provides a world manifold with a spinor structure
\cite{3sar,sard10}.

Given
a Minkowski space $M$ with the Minkowski metric $\eta$, let
$\Bbb C_{1,3}$ be the complex Clifford algebra generated by elements
of $M$.
A spinor space $V$ is defined to be a minimal left ideal of $\Bbb C_{1,3}$ on
which this algebra acts on the left. We have the representation
\begin{equation}
\g: M\otimes V \to V \label{521}
\end{equation}
of elements of the Minkowski space $M$ by
Dirac's matrices $\g$ on $V$.

Let us consider a bundle of complex Clifford algebras $\Bbb C_{3,1}$
over $X^4$. Its subbundles are both a spinor bundle $S_M\to X^4$ and the
bundle $Y_M\to X^4$ of Minkowski spaces of generating elements of
$\Bbb C_{3,1}$.
To describe Dirac fermion fields on a world manifold, one must
require that $Y_M$ is isomorphic to the cotangent bundle $T^*X$
of a world manifold $X^4$. It takes place if the structure group $GL_4$
of $LX$ is reducible to the Lorentz group
$L=SO(3,1)$ and $LX$
contains a reduced $L$ subbundle $L^hX$ such that
\begin{equation}
Y_M=M^hX=(L^hX\times M)/L.\label{10.2}
\end{equation}
In this case, the spinor bundle $S_M$ is associated with the $L_s$-lift
$P_h$ of $L^hX$:
\begin{equation}
S_M=S_h=(P_h\times V)/L_s.\label{510}
\end{equation}
For the sake of simplicity, we shall identify $P_h$ with $L^hX$.

In accordance with the well-known theorem, there is the 1:1 correspondence
between the reduced subbubdles $L^hX$ of $LX$ and
the tetrad gravitational fields $h$ identified with global sections
of the quotient bundle
\begin{equation}
\Si:= LX/L\to X^4. \label{10.7}
\end{equation}
This bundle is the 2-fold cover of the bundle of pseudo-Riemannian
bilinear forms in cotangent spaces to $X^4$. Global sections of the
latter are pseudo-Riemannian metrics $g$ on $X^4$.

Given a tetrad field $h$, let $\Psi^h$ be an atlas of
$LX$ such that the corresponding local sections $z_\xi^h$ of $LX$
take their values into the reduced subbundle $L^hX$.
With respect to an atlas $\Psi^h$ and a
holonomic atlas $\Psi^T=\{\psi_\xi^T\}$ of $LX$, the tetrad field $h$
can be represented by a family of $GL_4$-valued tetrad functions
\be
&& h_\xi=\psi^T_\xi\circ z^h_\xi,\\
&&dx^\la= h^\la_a(x)h^a,
\ee
which carry atlas (gauge) transformations between fibre bases
$\{dx^\la\}$ and $\{h^a\}$ of $T^*X$ associated with $\Psi^T$ and $\Psi^h$
respectively. The well-known relation
\begin{equation}
g^{\mu\nu}=h^\mu_ah^\nu_b\eta^{ab} \label{9}
\end{equation}
takes place.

Given a tetrad field $h$, one can define the representation
\begin{equation}
\g_h: T^*X\otimes S_h=(L^h\times (M\otimes V))/L\to (L^h\times
\g(M\otimes V))/L=S_h, \label{L4}
\end{equation}
\[
\wh dx^\la=\g_h(dx^\la)=h^\la_a(x)\g^a,
\]
of cotangent vectors to a world manifold $X^4$ by Dirac's $\g$-matrices
on elements of the spinor bundle $S_h$.

Let $A_h$ be a connection on $S_h$ associatede with a principal
connection on $L^hX$ and $D$
the corresponding covariant differential. Given the
representation (\ref{L4}), one can construct the Dirac operator
\[
\cD_h=\g_h\circ D: J^1S_h\to T^*X\op\otimes_{S_h}VS_h\to VS_h
\]
on $S_h$. Then, we can say that sections of
the spinor bundle $S_h$ describe Dirac fermion fields in the presence of
the tetrad gravitational field $h$.

The crucial point consists in the fact that,
for different tetrad fields $h$ and $h'$, Dirac fermion fields are
described by sections of spinor bundles associated
with different reduced $L$-principal subbundles
of $LX$ and so, the representations $\g_h$ and $\g_{h'}$
(\ref{L4}) are not equivalent.

It follows that
Dirac fermion field must be regarded only in a pair with a certain
tetrad gravitational field $h$. These pairs are represented by sections
of the composite spinor bundle
\begin{equation}
S\to\Si\to X^4 \label{L1}
\end{equation}
where $S\to\Si$ is a spinor bundle associated with the $L$ principal
bundle $LX\to\Si$ \cite{3sar,sard10}. In particular, every spinor bundle
$S_h$ (\ref{510}) is isomorphic to restriction of $S$ to $h(X^4)\subset\Si$.

Since, for different tetrad fields $h$ and $h'$,
the representations $\gamma_h$ and $\gamma_{h'}$
(\ref{L4}) are not equivalent,
even weak gravitational fields, unlike
matter fields and gauge potentials, fail to form an affine space
modelled on a linear space of deviations of some background field.
They thereby do not satisfy the superposition principle and can not be
quantized by usual methods, for in accordance with the algebraic
quantum field theory quantized fields must constitute a linear space.
This is the common feature of Higgs fields. In algebraic quantum field
theory, different Higgs fields correspond to nonequivalent Gaussian
states of a quantum field algebra. Quantized deviations of a Higgs
field can not change a state of this algebra and so, they fail to
generate a new Higgs field.

At the same time, one can examine
superposable deviations $\si$ of a tetrad gravitational field $h$ so
that $h+\si$ is not a tetrad gravitational field \cite{sard90,3sar}.
In the coordinate form, these deviations read
\ben
&&\wt h^\mu_a=s^b{}_ah^\mu_b = (\dl^b_a+\si^b{}_a)h^\mu_b=
s^\mu{}_\nu h^\nu_a =(\dl^\mu_\nu+\si^\mu{}_\nu)h^\nu_a=h^\mu_a+
\si^\mu{}_a, \label{3.5}\\
&&\wt h^a_\mu=g_{\mu\nu}\eta^{ab}\wt h^\nu_b=s_\mu{}^\al h^a_\al,
\nonumber\\
&&\wt h^\mu_a\wt h^a_\nu\neq \dl^\mu_\nu,\qquad \wt h^\mu_a\wt
h^b_\mu\neq \dl^b_a.\nonumber
\een
Note that the similar factors have been investigated by R.Percacci
\cite{per2,per}.

In bundle terms, we can describe the deviations (\ref{3.5}) as
the special morphism $\Phi_2$ of the cotangent bundle.
Given a gravitational field $h$ and the corresponding representation
morphism $\g_h$ (\ref{L4}), the morphism $\Phi_2$ yields another
$\g$-matrix representation
\ben
&&\wt\g_h=\g_h\circ\Phi_2,\label{10.14}\\
&&\wt\g_h(h^a)=s^a{}_b\g_h(h^b)=s^a{}_b\g^b,\nonumber
\een
of cotangent vectors, but on the same spinor bundle $S_h$. Therefore,
deviations (\ref{3.5}) and their superposition $\si + \si'$ can be defined.

Let us note that, to construct a Lagrangian density of deviations
$\epsilon$ of a
gravitational field, one usually utilize a familiar Lagrangian
density of a gravitationsl field $h' = h + \epsilon$
where $h$ is treated as a background field. In case of the
deviations (\ref{3.5}), one can not follow this method, for quantities
$\wt h$ fail to be true tetrad fields. To overcome this difficulty, we
use the fact that
the morphisms $\Phi_2$ appears also in the dislocation
gauge theory of the translation group.  We therefore may
apply the Lagrangian densities of this theory in order to describe
deviations $\si$ (\ref{3.5}).
They differ from the familiar gravitational
Lagrangian densities. In particular, they contain the
mass-like term.
Solutions of the corresponding field equations show that fields $\si$
make contribution to the standard gravitational effects. In particular,
they lead to the "Yukawa type" modification of Newton's gravitational
potential.

Note that a world manifold $X^4$ must satisfy
the well-known global topological conditions in order that gravitational
fields, space-time structure and spinor structure can exist. To
summarize these conditions, we assume that $X^4$ is not compact and
the linear frame bundle $LX$ over $X^4$ is trivial.

\section{Deviations of tetrad fields}

Let $\pi_P:P\to X$ be a principal bundle with a structure
Lie group $G$ which acts freely and
transitively on $P$ on the right:
\[
r_g : p\mapsto pg, \qquad
p\in P,\quad g\in G.
\]
Note that a principal bundle $P$ is also the general affine bundle modelled on
the associated group bundle $\wt\pi:\wt P\to X$ with the standard fibre
$G$ on which the structure group $G$ acts by the adjoint representation.
The corresponding bundle morphism reads
\[
\wt P\times P\ni (\wt p,p)\mapsto \wt pp\in P.
\]

Let $K$ be a closed subgroup of $G$. We have the composite manifold
\begin{equation}
\pi_{\Si X}\circ\pi_{P\Si}: P\to P/K\to X \label{7.16}
\end{equation}
where
\[
P_\Si:=P\to P/K
\]
is a principal bundle with the structure group $K$ and
\[
\Si_K=P/K=(P\times G/K)/G
\]
is the $P$-associated bundle with the standard fiber $G/K$ on which the
structure group $G$ acts on the left.

Let the structure group $G$ be reducible to its closed subgroup $K$.
Recall the 1:1 correspondence
\[
\pi_{P\Si}(P_h)=(h\circ\pi_P)(P_h)
\]
between the global sections $h$ of the
bundle $P/K\to X$ and the reduced $K$-principal
subbundles $P_h$ of $P$ which consist with
restrictions of the principal bundle $P_\Si$ to $h(X)$.

Let us consider the composite manifold
\begin{equation}
Y=(P\times V)/K\to P/K\to X \label{7.19}
\end{equation}
where the bundle
\[
Y_\Si:= (P\times V)/K\to P/K
\]
is associated with the $K$-principal bundle
$P_\Si$. Given a reduced subbundle $P_h$ of $P$,
the associated bundle
\[
Y_h=(P_h\times V)/K
\]
is isomorphic to the restriction of $Y_\Si$ to $h(X)\subset \Si_K$.

Note that the manifold $(P\times V)/K$ possesses also the structure of the
bundle
\begin{equation}
Y=(P\times (G\times V)/K)/G \label{10.3}
\end{equation}
associated with the principal bundle $P$. Its standard fibre
is $(G\times V)/K$ on which the structure group $G$ of $P$ (and its
subgroup $K$) acts by the law
\[
G\ni g: (G\times V)/K\to (gG\times V)/K.
\]
It differs from action of the structure group $K$ of $P_\Si$ on this
standard fibre. As a
shorthand, we can write the latter in the form
\[
K\ni g: (G\times V)/K\to (G\times gV)/K.
\]
However, this action fails to be canonical and depends on existence and
specification of a global section of the bundle $G\to G/K$.
If the standard fibre $V$ of the bundle $Y_\Si$ carriers representation
of the whole group $G$, these two actions are equivalent, otherwise
in general case.

Let $\Phi$ be an isomorphism of the principal bundle $P$ over $\Id X$. It is
expressed as
\begin{equation}
\Phi(p)= pf_s(p), \qquad p\in P,\label{10.4}
\end{equation}
where $f_s$ is a $G$-valued equivariant function
\[
f_s(pg)=g^{-1}f(p)g, \qquad g\in G,
\]
on $P$. There is the 1:1 correspondence
\[
s(\wt\pi(p))p=pf_s(p)
\]
between such functions and global sections $s$ of the corresponding
group bundle $\wt P$.

Every principal isomorphism $\Phi$ (\ref{10.4}) yields the following
morphism of the composite manifold (\ref{7.19}):
\begin{equation}
\Phi_1: (p\times v)/K\mapsto (pf_s(p)\times v)/K. \label{10.5}
\end{equation}
It is the morphism of $Y$ as the $P$-associated bundle (\ref{10.3}).
If the standard fibre $V$ of the bundle $Y_\Si$ admits a
representation of the whole group $G$, every principal isomorphism
$\Phi$ (\ref{10.4}) of $P$ generates another morphism of the composite
manifold (\ref{7.19}):
\begin{equation}
\Phi_2: (p\times v)/K\mapsto (p\times f_s(p)v)/K. \label{10.6}
\end{equation}
In comparison with (\ref{10.5}), this is a morphism over $\Id\Si$.
If the function $f_s$ is $K$-valued, the morphisms (\ref{10.5}) and
(\ref{10.6}) consist with each other and come to a familiar gauge
morphism of the bundle $Y_\Si$.

In gravitation theory, we have the composite manifold
\begin{equation}
LX\to\Si\to X^4 \label{10.8}
\end{equation}
where $\Si$ is the quotient bundle (\ref{10.7}), the associated composite
spinor bundle $S$ (\ref{L1}) and the composite bundle
\begin{equation}
MX:=(LX\times M)/L\to \Si\to X^4 \label{10.9}
\end{equation}
of Minkowski spaces.

Every principal isomorphism $\Phi$ of the linear frame
bundle $LX$ yields the morphisms $\Phi_1$ (\ref{10.5}) of the composite
bundles $S$ and $MX$ and the morphism $\Phi_2$ (\ref{10.6}) of the composite
bundle MX.

Let $h$ be a section of the the quotient bundle $\Si$ (\ref{10.7}).
A principal isomorphism $\Phi$  (\ref{10.4}) of $LX$ sends the reduced
principal bundle $L^hX$ to some reduced principal bundle $L^{h'}X$. In
other words, it transforms the tetrad field $h$ to the tetrad field
\[
h'(x)= (\pi_{P\Si}\circ \Phi)(h^{-1}(x)).
\]
The corresponding morphisms $\Phi_1$ of the composite bundles
$S$ and $MX$ determines the bundle morphisms
\begin{equation}
\Phi_1: S_h\to S_{h'}, \qquad \Phi_1: M^hX\to M^{h'}X \label{10.10}
\end{equation}
so that
\[
\g_{h'}\circ\Phi_1=\Phi_1\circ\g_h
\]
where $\g_h$ and $\g_{h'}$ are the representations (\ref{L4}).
Given an atlas $\{z^h_\xi\}$ of the reduced principal bundle $L^hX$,
let us provide $L^{h'}X$ and associated bundles with the atlas
\begin{equation}
z^{h'}_\xi(x)=z^h_\xi(x)f_s(z^h_\xi(x)). \label{10.16}
\end{equation}
With respect to these atlases, the morphisms (\ref{10.10}) read
\begin{equation}
\Phi_1(h^a)=h'^{\wt a}, \qquad \Phi_1(v_A(x))=v'^{\wt A}(x) \label{10.11}
\end{equation}
where $\{h^a\}$, $\{v_A(x)\}$, $\{h'^{\wt a}\}$ and $\{v'_{\wt A}(x)\}$ are
the corresponding bases of $M^hX$, $S_h$, $M^{h'}X$ and $S_{h'}$
respectively.

It should be noted that the bundles $M^hX$ and $M^{h'}X$ (\ref{10.2})
are isomorphic to the same cotangent bundle $T^*X$, but provided with
different Minkowski structures. Therefore, $\Phi_1$ (\ref{10.10}) is an
isomorphism of the cotangent bundle $T^*X$. If $h'\neq h$, there is
however no isomorphism of the spinor bundle $S_h$ so that the
representations $\g_h$ and $\g_{h'}$ would be equivalent.

Let $h$ be a section of the the quotient bundle $\Si$ and $\Phi$
a principal isomorphism of $LX$. In contrast with $\Phi_1$, the
corresponding morphism $\Phi_2$ (\ref{10.6}) determines the morphism of
$M^hX$ to itself:
\begin{equation}
\Phi_2:(p\times  e^a)/L\mapsto(p\times f_s(p)(e^a))/L, \qquad p\in
L^hX,  \label{10.13}
\end{equation}
where $\{e^a\}$ is the basis of the Minkoski space $M$. Its coordinate
expression relative to an atlas $\{z^h_\xi\}$ is exactly (\ref{3.5}).
The morphism (\ref{10.13}) does not alter the tetrad field $h$, but transforms
the cotangent bundle $T^*X$ of a world manifold. We call it the
deformation of the cotangent bundle. It is readily observed that, whenever
$h$, there is the 1:1 correspondence between
these deformations and the section of the group bundle $\wt LX$.

Since deformations (\ref{10.13}) transform the cotangent bundle, but
not the spinor bundle $S_h$, one can say that they violate the
correlation between the Lorentz structure and the spinor
structure on a world manifold. As a consequense, the deformation
(\ref{10.13}) yields another $\g$-matrix representation
of cotangent vectors to a world manifold on the spinor bundle $S_h$:
\begin{equation}
\wt\g_h=\g_h\circ\Phi_2: (p\times(e^a\otimes v_A))/L\mapsto
(p\times\g(f(p)e^a\otimes v_A))/L, \qquad p\in L^hX, \label{10.15}
\end{equation}
where $\{v_A\}$ is a basis of the standard fibre $V$.
The coordinate form of this representation is given by the expression
(\ref{10.14}).

Thus, we can model the nongravitational deviations
(\ref{3.5}) of a gravitational field by the deformations (\ref{10.13})
of the cotangent bundle. Since the representations $\wt\g_h$
and $\g_h$ are defined on the same spinor bundle $S_h$, these
deviations exist
\[
s^a{}_b=\dl^a_b + \si^a{}_b
\]
and their superposition $\si + \si'$ can be defined.

In particular, the Dirac operator corresponding to the representation
$\wt\g_h$ takes the form
\begin{equation}
\wt{\cD}=\wt\g_h(dx^\mu)D_\mu\phi_h=h^\mu_a(x)s^a{}_b(x)\g^b D_\mu\phi_h
= h^\mu_a(x)\g^as^\nu{}_\mu(x)D_\nu\phi.\label{3.6}
\end{equation}
on sections $\phi_h$ of the spinor bundle $S_h$.

It should be noted that, given a holonomic atlas of $LX$, it is the
function $s^\nu{}_\mu(x)$ which
does not depend on a gravitational field, that is, the tetrad functions
$h^\mu_a$ and the deviations $\si^\mu{}_\nu$ are independent dynamic
variables.

Recall that, if the function $f$ which determines the principal
morphism $\Phi$ is $L$-valued, the representations $\wt\g_h$ and
$\g_h$ are isomorphic. For an infinitesimal element $\si$,
we then have $\si_{ab}= -\si_{ba}.$

Let us remark that the morphisms $\Phi_1$ and $\Phi_2$ are the equivalent
transformations
of the cotangent bundle regarded as the $GL_4$-bundle. Therefore, if
world symmetries are  not  broken  (e.g.,  there  are  no  fermion
fields), the bundle $T^*X$ "loses"  the Lorentze structure and
the transmutations
\be
&&M^h_xX=(p\times f(p)M)/L= (p\times f(p)T^*)/G\\
&&\qquad=(pf(p)\times T^*)/G\to (pf(p)\times M)/L=M^{h'}_xX
\ee
of deviations $\si$ of a gravitational field $h$ into a new
gravitational field $h'$ may take place. Relative to the atlas (\ref{10.16}),
these transmutations take the coordinate form
\[
h'^{\wt a}_\mu=s_b{}^ah^b_\mu=\wt h^a_\mu,\qquad h'^\mu_{\wt a}\neq \wt
h^\mu_a.
\]

\section{Deviations of metric fields}

Without regard to fermion fields, one can choose metric functions
$g^{\mu\nu}$ as gravitational variables and examine their small
deviations (\ref{2.6}). However, if a space-time decomposition is
considered, these deviations also fail to form a linear space in general.

Recall that, in virtue of the well-known theorems,
if the structure group of $LX$ is reducible to the structure Lorentz group,
the latter, in turn, is reducible to its maximal compact subgroup $SO(3)$.
It follows that, for every reduced subbundle
$L^hX$, there exist a reduced subbundle $FX$ of $LX$ with the
structure group $SO(3)$ and the corresponding (3+1) space-time decomposition
\[
TX = FX\oplus T^0X
\]
of the tangent bundle of $X^4$ into a 3-dimensional
spatial distribution $FX$ and its time-like orthocomplement $T^0X$.
There is the 1:1 correspondence
\[
FX\rfloor \Omega = 0
\]
between the nonvanishing 1-forms $\Omega$ on a manifold $X$ and
the 1-codimensional distributions on $X$.
Then, we get the following modification of the well-known theorem
\cite{3sar,1sard}.
\begin{itemize}
\item For every gravitational field $g$ on a world  manifold
$X^4$, there exists an associated pair $(FX,g^R)$ of a
space-time distribution $FX$ generated by a tetrad 1-form
\[
h^0=h^0_\mu dx^\mu
\]
and a Riemannian metric $g^R$, so that
\begin{equation}
g^R=2h^0\otimes h^0-g=h^0\otimes h^0 +k \label{2.1}
\end{equation}
where $k$ is the Riemannian metric in the subbundle $FX$.
Conversely,
given a Riemannian metric $g^R$,  every  oriented  smooth  3-dimensional
distribution $FX$ with a generating form $\Omega$ is a space-time
distribution compatible with the gravitational field $g$ given by
expression (\ref{2.1}) where
\[
h^0=\frac{\Omega}{|\Omega|},\qquad |\Omega|^2=g^R(\Omega,\Omega) =
g(\Omega,\Omega).
\]
\end{itemize}

The triple $(g,FX,g^R)$ (\ref{2.1})
sets up uniquely a space-time structure on a
world manifold.
The Riemannian metric $g^R$ in the triple (\ref{2.1}) defines a
$g$-compatible distance function on a
world manifold $X^4$.  Such a function brings $X^4$ into a metric space whose
locally Euclidean topology is equivalent to the manifold topology on $X^4$.

Given a gravitational field $g$ and a $g$-compatible space-time
distribution $FX$, let $k$ be a spatial part of the world metric $g$.
If a world metric $g'$ results from some linear deviation
\[
g' = g - \e
\]
of $g$, one can require the spatial parts $k'$ of $g'$ to be a linear
deviation
\[
k' = k + \e_k
\]
of $k$. It takes place if there exists a space-time distribution $FX$
compatible with both $g$ and $g'$. In this case, we have
\be
&&k'  = k + \e+\frac{\Omega\otimes \Omega}{|g(\Omega,\Omega)|^2}
\e(\Omega,\Omega),\\
&&\e(\Omega,\Omega) = \e^{\al\bt}\Omega_\al
\Omega_\bt,
\ee
where $\Omega$ is a generating form of the distribution $FX$. For instance,
given a triple $(g, FX, g^R)$, every linear deviation
\[
g'^R=g^R-\e^R
\]
of the Riemannian metric $g^R$ in this triple involves the
linear deviation
\[
g' = g + \e^R-2\Omega\otimes \Omega\frac{\e^R(\Omega,
\Omega)}{|g^R(\Omega,\Omega)|^2}
\]
of the pseudo-Riemannian metric $g$ in and its spatial part
\[
k' = k - \e^R-\Omega\otimes \Omega\frac{\e^R(\Omega,
\Omega)}{|g^R(\Omega,\Omega)|^2}
\]
so that the triple $(g',FX,g'^R)$ is associated with the same
distribution $FX$.

Obviously, there are pseudo-Riemannian metrics $g$ and $g'$ which fail to
admit the same space-time distribution. Their superposition is accompanied
by superposition of space-time distributions which we face, e.g., in the
case of gravitational singularities of the caustic type \cite{3sar,1sard}.

The deviations (\ref{3.5}) also yields the corresponding
nongravitational deviations of a metric field:
\ben
&&\wt g^{\mu\nu}=\wt h^\mu_a\wt h^\nu_b\eta^{ab}=s^\mu{}_\al
s^\nu{}_\bt g^{\al\bt},\nonumber\\
&&\wt g_{\mu\nu}=\wt h^a_\mu\wt h^b_\nu\eta_{ab}=s_\mu{}^\al
s_\nu{}^\bt g_{\al\bt},\label{3.7}\\
&&\wt g^{\mu\nu}\wt g_{\mu\al}\neq\dl^\nu_\al. \nonumber
\een
The quantity $\wt g$ in this expression is not a world metric.
In comparison with the relation (\ref{2.6}), we have
\be
&&\wt g^{\mu\nu}\approx g^{\mu\nu}+\si^{\mu\nu},\\
&&\wt g_{\mu\nu}\approx g_{\mu\nu}+g_{\mu\al}g_{\nu\bt}
\si^{\al\bt},
\ee
for small deviations
\[
\si^{\mu\nu}=\sigma^a{}_b h^\mu_a h^{b\nu}.
\]

\section{Dislocated manifolds}

The deformations morphisms (\ref{10.13}) of the cotangent bundle appear
in the gauge theory of the translation group \cite{sard90,3sar}.

Let the tangent bundle $TX$ be provided
with the canonical structure of the affine tangent bundle. It is
coordinatized by $(x^\mu, u^\la)$ where $u^\al\neq \dot x^\al$ are the
affine coordinates.

Every affine connection $A$ on $TX$ is brought into the sum
\begin{equation}
A=\G+\si=dx^\mu\otimes[
\frac\dr{\dr x^\mu}+
(\G^\al{}_{\bt\mu}u^\bt+\si^\al{}_\mu)\frac\dr{\dr u^\al}] \label{4.4}
\end{equation}
of a linear connection $\G$ and a soldering form
\[
\si=\si^\la{}_\mu(x)dx^\mu\otimes\frac\dr{\dr u^\la}
\]
which plays the role of a gauge translation potential.

In the conventional gauge theory of the affine  group, one
faces the problem of physical interpretation of both gauge
translation potentials and sections $u(x)$ of the affine tangent
bundle $TX$. In field theory, no fields possess the transformation
law
\[u(x)\to u(x) + a\]
under the Poincar\'e translations.

At the same time, one observes such fields in the gauge theory
of dislocations \cite{kad} which is based on the fact that, in the presence
of dislocations, displacement vectors $u^k,\, k = 1,2,3,$
of small deformations are  determined  only  with  accuracy  to  gauge
translations
\[
u^k\to u^k+a^k(x).
\]
In this theory, gauge translation potentials $\si^k{}_i$ describe
the plastic distortion, the covariant derivatives
\[
D_iu^k =\dr_i u^k - \si^k{}_i
\]
consist with the elastic distortion, and the
strength
\[\cF^k{}_{ij}=\dr_i \si^k{}_j -\dr_j \si^k{}_i\]
is the dislocation density.
Equations of the dislocation theory
are derived from the gauge invariant Lagrangian density
\begin{equation}
\cL=\mu D_iu^kD^iu_k+\frac{\la}{2} D_iu^iD_mu^m-\epsilon \cF^k{}_{ij}
\cF_k{}^{ij} \label{4.7}
\end{equation}
where $\mu$ and $\lambda$ are the Lame coefficients of isotropic media.
These equations however are not
independent of each other since a displacement field
$u^k(x)$ can be removed by gauge translations and, thereby, it fails to be a
dynamic variable.

In the spirit of the gauge dislocation theory, it was suggested that
the gauge potentials of the Poincar\'e translations can describe new
geometric structure ({\it sui generis} dislocations) of a world
manifold \cite{ede,sard87}.

Let the tangent bundle $TX$ be provided with an affine
connection (\ref{4.4}). We consider the following two morphisms:
\begin{itemize}
\item
the morphism
\[
\wh o: TX\to TTX
\]
which is the morphism defined by the connection $A$ and restricted to
the global zero section $\wh 0$ of $TX$, that is,
\[
\wh o=A\circ\wh 0: \wh 0(X)\op\times_XTX\to TTX;
\]
\item
the geodesic morphism of $TX$ onto $X$:
\[
\zeta: TX\ni u\to \zeta(x,u,1)\in X,\qquad x=\pi_X(u),
\]
where $\zeta(x,u,s)$ is the geodesic defined by the linear part
$\Gamma$  of the affine connection (\ref{4.4}) through the point $x$ in the
direction $u$. We shall call $\si$ the deformation field.
\end{itemize}

By dislocation of a world manifold $X$, we call the following
bundle morphism over $\Id X$:
\ben
&&\rho=T\zeta\circ\wh o\colon\ TX\to TX, \label{4.8}\\
&&\rho: \frac\dr{\dr x^\mu}\to \frac\dr{\dr x^\mu}+
(\G^\al{}_{\bt\mu}u^\bt+\si^\al{}_\mu)\frac\dr{\dr u^\al}
\to(\dl^\al_{\mu}+\si^\al{}_\mu)
\frac\dr{\dr x^\alpha}=s^\al{}_\mu\frac\dr{\dr x^\al}.\nonumber
\een
Here, we use the relations
\be
&&\zeta^\mu(x,\la u,1)=\zeta^\mu(x,u,\la),\qquad \la\in \Bbb R,\\
&&\frac\dr{\dr u^\al}\zeta^\mu(x,u,1)|_{u=0}=\dl^\mu_\al,
\ee
and the expression
\[
D_\mu u^\al|_{u=0}=(\dr_\mu u^\al+\G^\al{}_{\bt\mu}
u^\bt+\si^\al{}_\mu)|_{u=0}=\si^\al{}_\mu
\]
for the covariant derivatives of a displacement field $u$.

Let $Y\to X$ be a fibred manifold and $J^1Y$ the jet manifold of $Y$.
The dislocation (\ref{4.8}) gives rise to the morphism
\[
J\rho: (x^\la,y^i,y^i_\la)\to (x^\la,y^i,s^\al{}_\la y^i_\al)
\]
of $J^1Y$ over $\Id Y$.

To define fields on a dislocated manifold, we
therefore can replace sections $w(x)$ of $J^1Y\to X$ and $w(y)$ of
$J^Y\to Y$ by sections
\[
\wt w(x)=(J\rho\circ w)(x),\qquad \wt w(y)=(J\rho\circ w)(y).
\]
If $\f$ is a section of the bundle $Y$, we have
\be
&&\wt{J^1\f}=J\rho\circ J^1\f,\\
&&(\wt{J^1\f})^i_\la=s^\al{}_\la(x) \dr_\al \f^i(x).
\ee
Let $\G$ be a connection on $Y$ and $D$ be the corresponding covariant
differential. On a dislocated manifold, we get
\be
&&\wt \G^i_\la(y)=s^\al{}_\la\G^i_\al(y)\\
&&\wt D=dx^\la\otimes \wt D_\la=
dx^\la\otimes s^\al{}_\la(x)(\dr_\al - \G^i_\al(y)\dr_i).
\ee

For instance, the Dirac operator on a dislocated manifold
takes the form
\[
\wt L_D=\g_h(dx^\la)\otimes\wt D_\la=
s^\al{}_\la\g_h(dx^\la)\otimes D_\al.
\]
This operator looks like the Dirac operator (\ref{3.6}) in the presence of
deviations (\ref{3.5}) of a tetrad gravitational field if the
morphism (\ref{3.6}) consists with the dual to
the morphism (\ref{4.8}). We therefore can apply Lagrangians of the field
theory on dislocated manifolds to deviations (\ref{3.5}).

A Lagrangian density of a scalar field $\phi$ on the dislocated
manifold reads
\[
\cL_{(m)}=\frac12(g^{\mu\nu}s^\al{}_\mu s^\bt{}_\nu D_\al\phi
D_\bt\phi-m^2\phi^2)\sqrt{-g}.
\]
Lagrangian  densities  $\cL_{(g)}$ of the gravity and $\cL_{(A)}$ of gauge
potentials are constructed by means of the modified curvature
\[\wt R^{ab}_{\mu\nu}=s^\e{}_\mu s^\bt{}_\nu
R^{ab}{}_{\e\bt}\]
and the modified strength
\be
&&\wt \cF=\wt\rho\circ \cF\circ J^1A,\\
&&\wt \cF^m_{\mu\nu}=s^\al{}_\mu s^\bt{}_\nu \cF^m_{\al\bt}.
\ee

The action functional and equations of motion of a point mass $m_0$
on the deformed manifold are given by expressions
\be
&&S=-m_0\int(g_{\al\bt}s^\al{}_\mu s^\bt{}_\nu v^\mu
v^\nu)^{1/2}ds, \nonumber\\
&&\frac{dv^\mu}{ds} + \wt \G^\mu{}_{\al\bt}v^\al v^\bt=0 \label{4.9}
\ee
where $v^\mu$ is the 4-velocity and the quantities $\wt\G$ look like
the Christoffel symbols of the "metric"
\[\wt g_{\mu\nu}=s^\al{}_\mu s^\bt{}_\nu g_{\al\bt},\]
but the interval $ds$ is defined by the true world metric $g$.

Let us note that, on the dislocated manifold, a world metric and the volume
form remain unchanged.

\section{Gauge theory of deformation fields}

The Lagrangian density $\cL_{(\si)}$ of translation gauge  potentials
$\si^\e{}_\mu$ can not be built in the Yang-Mills form because the
Lie algebra of the affine group does not admit an invariant nondegenerate
bilinear form. To construct $\cL_{(\si)}$, one can utilize the torsion
\[
\cF^\al{}_{\nu\mu}=D_\nu\si^\al{}_\mu  -D_\mu\si^\al{}_\nu
\]
of the connection $\G$ with respect to the soldering form $\si$.

The general form of a Lagrangian density $\cL_{(\si)}$
is given by the expression
\be
&&\cL_{(\si)}=\frac12[a_1\cF^\mu{}_{\nu\mu} \cF_\al{}^{\nu\al}+
a_2\cF_{\mu\nu\si}\cF^{\mu\nu\si}+a_3\cF_{\mu\nu\si}\cF^{\nu\mu\si}\\
&&\qquad +a_4\e^{\mu\nu\si\g}\cF^\e{}_{\mu\e}
\cF_{\g\nu\si}-\mu\si^\mu{}_\nu\si^\nu{}_\mu+
\la\si^\mu{}_\mu \si^\nu{}_\nu]\sqrt{-g}
\ee
where $\e^{\mu\nu\si\g}$ is the Levi-Civita tensor.

The mass-like term in $\cL_{(\si)}$ is originated from the Lagrangian
density (\ref{4.7}) for displacement fields $u$ under the gauge condition
$u=0$.

It seems natural to require the component ${\bf t}^{00}_{(\si)}$
of a metric energy-momentum tensor of deformation fields $\si$
on the Minkowski space be positive.  This condition
implies the following constraints on the constants in $\cL_{(\si)}$:
\[a_4=0,\quad a_1\geq 0,\quad a_2\geq 0, \quad a_3+2a_2=0,\quad \mu\geq
0,\quad \lambda\leq \frac14\mu.\]
The Lagrangian density $\cL_{(\si)}$ then takes the form
\[\cL_{(\si)}=\frac12[a_1\cF^\mu{}_{\nu\mu}\cF_\al{}^{\nu\al}+a_2
\cF_{\mu\nu\si}(\cF^{\mu\nu\si}-2\cF^{\nu\mu\si})-\mu\si^\mu{}_\nu
\si^\nu{}_\mu+\la\si^\mu{}_\mu\si^\nu{}_\nu]\sqrt{-g}.\]
We here use the decomposition of the tensor $\cF^\la{}_{\mu\nu}$ in
three irreducible parts
\be
&&\cF^\la{}_{\mu\nu}=\wt \cF^\la{}_{\mu\nu}+\frac13(\dl^\la_\nu
\cF_\mu-\dl^\la_\mu\cF_\nu)+\e^\la{}_{\mu\nu\al}\wt\cF^\al,\\
&&\cF_\mu=\cF^\al{}_{\mu\al},\qquad \wt \cF^\al=\frac1{6}\e
^{\al\mu\nu\si}\cF_{\mu\nu\si},
\ee
where $\cF_\mu$ is the spur, $\wt \cF^\al$ is the pseudo-spur, and $\wt \cF$
is the spur-free part of the tensor $\cF$.

The total Lagrangian density includes Lagrangian densities $\cL_{(m)}$
of mater fields, $\cL_{(A)}$ of gauge potentials, and $\cL_{(g)}$ of a
gravitational field. Matter sources of a deformation field $\si$ then
are the following:
\begin{itemize}
\item
the short canonical energy-momentum tensor of matter fields
\[-\frac{\dl \cL_{(m)}}{\dl\si^{\mu\nu}}=-(s^{-1})_{\nu\bt}
D_\mu \phi\frac{\dr \cL_{(m)}}{\dr D_\bt\phi}=-(s^{-1})_{\nu\bt}
({\bf T}_{(m)}{}^\bt_\mu +\dl^\bt_\mu L_{(m)})\]
where ${\bf T}_{(m)}$ denotes a canonical energy-momentum tensor of matter
fields;
\item
the short metric energy-momentum tensor ${\bf t}_{(A)}$ of gauge potentials:
\[-\frac1{\e^2}a^G_{mn}\wt g^{\g\bt} s^\al{}_\mu
\cF^m_{\al\bt}\cF^n_{\g\bt}\sqrt{-g}\]
where $\wt g^{\g\bt}$ is the "metric" (\ref{3.7});
\item the curvature tensor
\[\kappa^{-1}g_{\nu\alpha}s^\e{}_\g
R^{\al\g}_{\mu\e}\sqrt{-g}\]
of a gravitational field.
\end{itemize}

Let us restrict ourselves to the case of a small field $\si$. We
neglect a gravitational field on the left-hand  side of equations for
$\si$ and keep only $\si$-free terms in matter sources.
Then, the Euler-Lagrange equations for a deformation field $\si$ read
\ben
&&\frac{\dl \cL_{(\si)}}{\dl\si^{\mu\nu}}=a_1(\eta_{\mu\nu}
\dr^\e \cF^\al{}_{\al\e}-\dr_\mu \cF^\al{}_{\al\nu})
+2a_2\dr^\e(\cF_{\mu\nu\e}-\cF_{\nu\mu\e}+
\cF_{\e\mu\nu})\nonumber\\
&&\qquad -\mu\si_{\mu\nu}+\la\eta_{\mu\nu}\si^\al{}_\al=
S_{\mu\nu},\nonumber\\
&& S_{\mu\nu}=-({\bf T}_{(m)\nu\mu}+g_{\nu\mu}\cL_{(m)})-\frac1{\e^2}
a^G_{mn} g^{\bt\g}\cF^m_{\mu\bt}\cF^n_{\nu\g}\sqrt{-g}
+\kappa^{-1}R_{\mu\nu}\sqrt{-g}.\label{4.10}
\een
One can replace the gravitation term in the equation (\ref{4.10}) by the
right-hand side of the Einstein equations. Equations for $\si$ then take
the form
\[
\frac{\dl\cL_{(\si)}}{\dl\si^{\mu\nu}}=({\bf t}_{(m)\nu\mu}-
{\bf T}_{(m)\nu\mu})-g_{\mu\nu}(\cL_{(m)}+\frac12{\bf t}_{(m)})-
g_{\mu\nu}\cL_{(A)}.
\]

The equation (\ref{4.10}) implies the equilibrium equation
\begin{equation}
\dr^\nu \frac{\dl\cL_{(\si)}}{\dl\si^{\mu\nu}}= -\mu
\dr^\nu\si_{\mu\nu}+\la\dr_\mu\si^\al{}_\alpha= \dr^\nu
S_{\mu\nu}.\label{4.11}
\end{equation}
Note that the right-hand side of this equation is equal neither
to zero nor to a gradient quantity in general. At the same time, this is a
pure gradient quantity if matter sources of the field $\si$ are gauge
potentials and scalar fields. These facts result in the important condition
\begin{equation}
\mu\neq 0,\qquad\mu\neq 4\la. \label{4.12}
\end{equation}

Since equations (\ref{4.10}) are linear, their solutions differ from each
other in solutions of the free field equations.
In the case of a free field $\si$, equation~(\ref{4.11}) reads
\[
-\mu\dr^\nu\si_{\mu\nu}+\la\dr_\mu\si^\al{}_\al = 0.
\]
Taking into account this relation, one can bring equations (\ref{4.10})
into the equations
\ben
&&4a_2\dr^\e(w_{\mu\e,\nu}+w_{\nu\mu,\e}-
w_{\nu\e,\mu}) + 2a_1(w^\al{}_{\nu,\mu\al}-w^\al{}_{\mu,
\nu\al})- \mu w_{\mu\nu} = 0, \label{4.13}\\
&&a_1\left[\frac{\la}{\mu}-1\right][\eta_{\mu\nu}\Box e-e_{,\mu\nu}] +
2a_1(w^\alpha{}_{\nu,\mu\al}+w^\al{}_{\mu,\nu\al})
-\mu e_{\mu\nu}+\la\eta_{\mu\nu}\si = 0 \label{4.14}
\een
where $\Box=\dr^\al\dr_\al$ and
\begin{equation}
e_{\mu\nu}=\frac12(\si_{\mu\nu}+
\si_{\nu\mu}),\quad w_{\mu\nu}=\frac12(\si_{\mu\nu}-
\si_{\nu\mu}), \quad e=\si^\al{}_\al. \label{4.15}
\end{equation}
It seems natural to choose the solution $w=0$ of equations (\ref{4.13}).
Equations (\ref{4.14}) then  can  be written in the form
\ben
&&e_{\mu\nu}=\frac{\mu-\la}{3\mu}(\eta_{\mu\nu}e-\frac{3a_1}{\mu}
e_{,\mu\nu}),\nonumber\\
&&\Box e+m^2e=0,\qquad m^2=\frac{\mu(\mu-4\la)}{3a_1(\mu-\la)},\label{4.16}
\een
where the quantity $m$ plays the role of a mass of deformation fields
$\si$. In virtue of the condition (\ref{4.12}), this mass is not equal to
zero.
Equations (\ref{4.16}) admit the following plane wave solutions
\[
e_{\mu\nu}=\frac{\mu-\la}{3\mu}\left[\eta_{\mu\nu}+
\frac{\mu-4\la}{\mu-\la}\frac{p_\mu p_\nu}{p^2}\right]a(p)e^{ipx},
\qquad p^2=m^2.
\]

Now, let us consider a model of a small deformation field $\si$ and a
small gravitational field
$g=\eta+2\e$
if their matter source is a motionless point mass $M$. In this
case, the right-hand side of equations (\ref{4.10}) reads
\[
-\frac12\eta_{\mu\nu}{\bf T}_{(m)}=-\frac12\eta_{\mu\nu}M\dl(r)
\]
where, by $(r,\phi,\theta)$, we denote spatial spherical coordinates.

Recalling the notations (\ref{4.15}), we can rewrite equations (\ref{4.10}) in
the form
\be
&&\frac{-a_1}{2}(e^\al{}_{\nu,\al\mu}-e^\al{}_{\mu,\al\nu})+
(4a_2+\frac{a_1}{2})(w^\al{}_{\mu,\al\nu}-w^\alpha{}_{\nu,
\al\mu})-4a_2\Box w_{\mu\nu}-\mu w_{\mu\nu}=0,\\
&&a_1[\eta_{\mu\nu}(e^{\al\e}{}_{,\al\e}-\Box e)-\frac12
(e^\al{}_{\nu,\al\mu}+e^\al{}_{\mu,\al\nu}+
w^\al{}_{\nu,\al\mu}+w^\al{}_{\mu,\al\nu})+e_{,\mu\nu}]\\
&&\qquad-\mu e_{\mu\nu}+\eta_{\mu\nu}\la
e=-\frac12\eta_{\mu\nu} M\dl(r).
\ee

These equations admit the static spherically symmetric solution with the
following nonzero components
\be
&&e_{rr}=-\frac1{\mu-\la}(3\la e_{00}+\frac12 M\dl(r)),\\
&&e_{\theta\theta}=-e_{00}r^2,\quad e_{\phi\phi}=-e_{00}r^2\sin^2
\theta,\\
&&\frac1{r^2}\frac{\dr}{\dr r}r^2\frac{\dr}{\dr r}e_{00}-m^2e_{00}=
-\frac16\frac\mu{a_1(\mu-\la)}M\delta(r),\\
&&e_{00}=\frac{\mu M}{24\pi a_1(\mu-\la)}\frac{e^{-mr}}{r}
\ee
where $m$ is the mass (\ref{4.16}).

Substituting  this  solution  into  equation (\ref{4.9}), we obtain the
modification of Newton's gravitational potential
\[
\wt \e=\e+e_{00}=-\frac{\kappa M}{8\pi r}\left(1-\frac{\kappa
^{-1}\mu}{3a_1(\mu-\la)}e^{-mr}\right).
\]
Such  a "Yukawa  type" modification of Newton's gravitational
potential is usually related to the hypothetical fifth fundamental force
\cite{sab}.

To contribute to standard gravitational effects, the fifth
interaction must be as universal as gravity. Its matter source  must
contain a mass or other parts of the energy-momentum
tensor.  This interaction must be described  by  a  massive  classical
field, though its mass is unusually small. A deformation field fits these
conditions.  For example, the mass (\ref{4.16}) is expressed by means of
constants of the Lagrangian density $\cL_{(\si)}$ where $\mu$ and
$\la$ make the sense of coefficients of "elasticity" of a space-time.

\end{document}